\documentclass[12pt,onecolumn,showpacs,amssymb,aps,nofootinbib,floatfix]{revtex4-1}
\usepackage{epsfig}

\newcommand{\eqcomma}{\phantom{AA},\phantom{AA}}

\newcommand{\order}[1]{ \mathcal{O} \left( #1 \right) }

\usepackage{color}

\begin{document}
\topmargin -0.8cm\oddsidemargin = -0.7cm\evensidemargin = -0.7cm
\preprint{}

\title{A quark-gluon plasma inspired model of the universe: Introduction and Inflation}
\author{Melissa Mendes$^{a,b}$, Giorgio Torrieri$^b$}
\affiliation{$\phantom{A}^a$ ITP, J. W. Goethe University, Frankfurt, Germany}
\affiliation{$\phantom{A}^{b}$IFGW, University of Campinas, Brazil}
\email{torrieri@g.unicamp.br}
\date{April 2018}

\begin{abstract}
We explain how a $SU(N_c)$ gauge theory, decoupled from the standard model and with a high-lying strong coupling scale, can incorporate apparently unrelated cosmological features, such as Inflation and dark matter, using well-understood dynamics from Quark-Gluon Plasma Physics.  In our scenario, the evolution of the universe is throughoutly hot: Inflation occurs due to the bulk viscosity peak during the mixed phase to deconfinement, while dark matter is composed of weakly interacting glueballs formed in the same phase. We parametrize the temperature dependence of the EoS and the viscosity expected from gauge theory, solve the Friedmann-Robertson-Walker (FRW) equations and compute the number of efoldings as a function of the free parameters of the model.
\end{abstract}

\maketitle
\section{Introduction}
Cosmology, the study of the origin of the universe \cite{cosmobook,weinberg,baumann}, has recently been characterized by an intriguing combination of phenomenological success and theoretical ambiguity.   The so-called $\Lambda$-CDM model has fitted most of the universe's macroscopic characteristics with just a few parameters, including gravitating but not otherwise interacting ``dark matter'' and an anomalously gravitating dark energy \cite{planck}.   The inflationary paradigm \cite{guth} has eliminated the necessity of fine-tuning to explain the global structure of spacetime, producing a nearly-homogeneous and nearly flat universe from generic initial conditions via an early exponentially expanding phase driven by a dynamically changing ``temporary cosmological constant''.
These parameters explain a wide variety of both present and past features, from galaxy rotation curves to structure formation.

However, at the moment, this phenomenological paradigm severely lacks a particle physics underpinning.  For instance, we have no idea of what the composition of dark matter is.  It has yet to be directly detected and models where it appears naturally, such as supersymmetry, have failed to be experimentally confirmed.  Within phenomenological models, dark matter is just assumed to be a dust of heavy but non-interacting particles, with residual interactions more and more tightly constrained experimentally \cite{constraint}.

Similarly, no one knows what is the nature of ``the inflaton''.  Its theoretical formulation is understood, at a semiclassical level, to be that of a nearly-flat (``slow-rolling'') false vacuum plateau, with a true vacuum where our universe moved after Inflation and stabilized after reheating.   No currently-known particle has the required characteristics for reproducing this behaviour and there is quite a lot of numerical evidence \cite{owe} that the quantum structure of such a theory is not well-defined.

This theoretical ambiguity requires new thinking, in particular whether different kinds of physics are capable of reproducing the same scenario. Here, natural candidates are non-abelian gauge theories. Unlike scalar field theories, there is little doubt about their fundamental mathematical and theoretical soundness \cite{wilson,creutz} and although their main qualitative features, notably confinement, are not rigorously derived from the Yang-Mills Lagrangian, there is a consensus regarding their basic nature. Crucially, this nature is ``universal'', that is, common to a family of theories which share basic properties such as asymptotic freedom.  Concurrently, the heavy ion program \cite{hicexp,hicexp2,hicexp3}  has done a great deal to elucidate the equilibrium and transport properties of these theories, on both a theoretical and a phenomenological level.

In this work, after a coincise overview of the thermal and transport properties of Yang-Mills theories, we shall argue that a Yang-Mills theory with a large number of colors ($N_c>3$) and no flavors \cite{thooft,panero}, with a strong-coupling scale of at least order \textit{TeV}, could provide a scenario to explain several independent features of standard cosmology.  We then focus on Inflation and try to see how it could emerge in such a model \cite{thesis}.

\section{A review of $SU(N_c)$ pure gauge theories}
\subsection{Basic theory and phase structure \label{phases}}
Confining pure Gauge theories (Gauge group $SU(N_c)$ with no fermions) have been extensively studied, as they provide a much simplified, both analytically and numerically but still qualitatively, similar model to $QCD$ \cite{thooft,creutz,panero}.   The Lagrangian of this theory is simply:

\begin{equation}
\label{lagrangian}
\mathcal{L} = -\frac{1}{4\lambda_{YM}(Q)} \mathrm{Tr}\left( F^{\mu \nu a} F^a_{\mu \nu}\right) \eqcomma F_{\mu \nu}^a = \partial_\mu A_\nu^a - \partial_\nu A_\mu^a + i \frac{\lambda_{YM}(Q)}{N_C} \sum_{b,c} f^{abc} A_\mu^b A_\nu^c
\end{equation}
where $A^\mu_a$ are gluon fields; $N_c$ stands for the free parameter number of colours;  $f^{ijk}$ are the structure constants associated with the $SU(N_c)$ group \cite{panero} and $\lambda_{YM}(Q)$ is a bare coupling constant defined at a momentum scale $Q$, in a thermal system, usually the temperature $T$.

There are two fundamental parameters of this theory, which for the purposes here can be thought of as independent.   The first is $N_c$, the number of colors, which specifies the gauge group and it is evident in the choice of Lagrangian.
The second parameter is absent in the classical Lagrangian but it appears when the theory is quantized as the scale at which the coupling of the theory becomes relevant, $\Lambda_{YM}$. Since $\lambda_{YM}(Q \rightarrow \infty) \rightarrow 0$, there must be a scale $\Lambda_{YM}$ for which $\lambda_{YM}(\Lambda_{YM}) \sim 1$.  While for QCD this scale is approximately $10^2$ MeV, generically it is an arbitrary parameter. Particularly, in the case where the $N_c \rightarrow \infty$ limit is reached continuously, $\Lambda_{YM}$ is independent of $N_c$ \cite{thooft}. This limit seems to be reached very fast for pure gauge theory \cite{panero}, although for theories with fermions its structure is more complicated \cite{menc}. In this work, we consider a case where $\Lambda_{YM} \sim TeV$.

The theory without fundamental fermions is characterized by a deconfinement transition \cite{kapusta,rafelski} between two phases: a plasma of $\order{N_c^2}$ massless gluons where the Lagrangian (\ref{lagrangian}) is manifest and a gas of massive glueballs \cite{glue1,glue2,glue3}. The lightest of those particles has a mass of the order of the phase transition temperature $(T_c)$, spin zero and no color dependence. In the 't Hooft limit \cite{thooft,panero} they are also weakly interacting, with a coupling proportional to $N_c^{-1}$. We note that physical $SU(3)$ QCD has a goldstone light mode, the pion, due to the presence of light quarks and, consequently, chiral symmetry. 
Therefore, the evidence against a light dark matter particle (``hot'' dark matter) means that we must assume the hidden gauge theory has no light flavors.

\begin{figure*}[h]
\epsfig{width=15cm,clip=1,figure=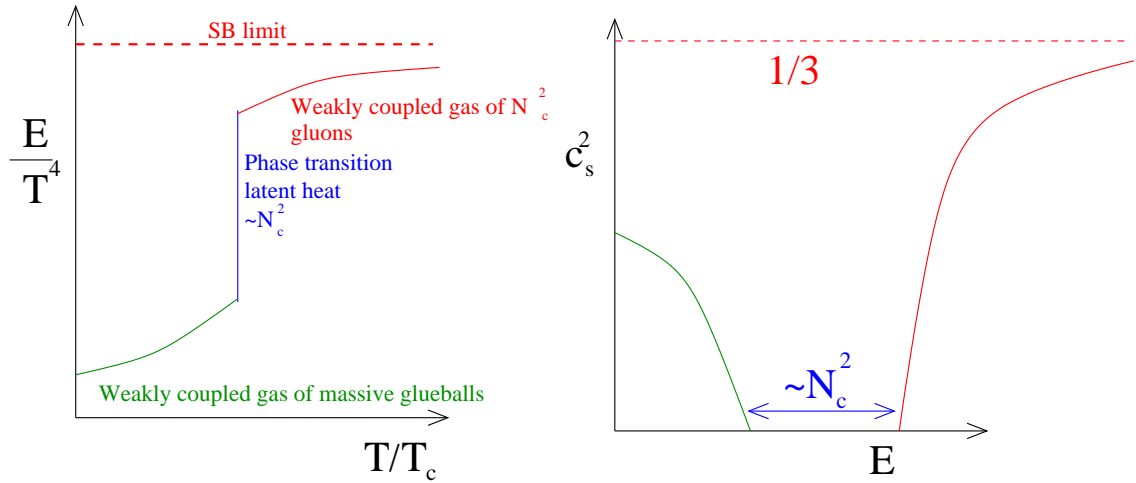}
\caption{\label{eos} Energy density (left panel) and speed of sound (right panel) as a function of temperature for Yang-Mills matter }
\end{figure*}

The full equation of state of the system is sketched in Fig. \ref{eos}. For pure gauge theory, the emergence of a $Z(N_c)$ theory, with the Polyakov loop as an expectation value \cite{polyakov}, ensures that at large $N_c$ the phase transition is of first-order. In-between around the critical temperature, the nature of the effective degrees of freedom is unclear - they may be quasiparticles \cite{peshier}, Hagedorn states \cite{jorge} or thermal excitations \cite{teaney} - but it is reasonable to suppose they interact strongly.

\begin{figure*}[h]
\epsfig{width=18cm,clip=1,figure=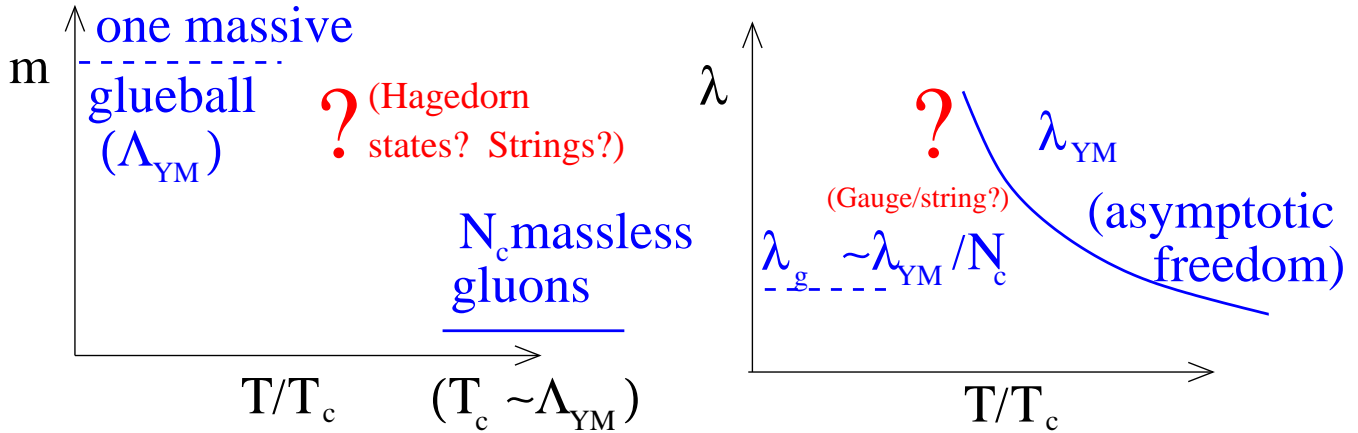}
\caption{\label{betafunc} As schematic representation  of the mass gap (left panel) and coupling constant (right panel) as a function of temperature for Yang-Mills matter }
\end{figure*}

The scale $\Lambda_{YM}$ determines, up to a factor $\order{1}$ calculable on the lattice \cite{creutz}, both the phase transition temperature $T_c$ and the mass of the confined low-lying state $m_h$. These two quantities can be determined logarithmically from the magnitude of the coupling constant at the renormalization UV scale, typically taken to be the Planck scale. In our theory, $\Lambda_{YM}$ also represents the scale of both Inflation and the formation of dark matter, hence it has to be of the order of \textit{TeV}, although still much smaller than the Planck scale, to ensure consistency with semiclassical gravity \cite{weakg}. Thus, neglecting higher spin Regge excitations, the effective Lagrangian is \cite{thooft,glue1,glue2,glue3}:

\begin{eqnarray}
\label{eftlagrangian}
\mathcal{L}_h  =  \frac{1}{2}\partial_{\mu}\phi\partial^{\mu}\phi - \frac{1}{2}m_h^{2}\phi^{2} - \sum_{n=3}^\infty \order{\frac{1}{N_c^n}} \left(\frac{p}{\Lambda_{ym}}\right)^n\phi^{n} +...
\end{eqnarray}

\noindent It represents weak interactions for low-temperature systems such as the present universe.

The equation of state of the system describes a massless gas at $T>T_c$ and a massive gas at $T<T_c$, both weakly interacting. The thermodynamics of both these systems is well-known \cite{rafelski}, such that at $T \gg T_c$, the EoS will be given by:
\begin{eqnarray}
p_g (T)= e_g(T)/3+ f_e (\lambda_{YM},T)-B  \\
e_g(T) \sim N_c^2 T^4 +  f_p (\lambda_{YM},T)+B  \\
B^2 \sim p_{g}(T_c) - p_{h}(T_c) \sim N_C^2 \Lambda_{YM}
\end{eqnarray}

\noindent{where} $p_g$ and $p_h$ stand respectively for the gluon and hadron, in this case the glueball, pressure and $e_g$ for the gluon energy density. B represents the bag constant (latent heat), $f_e(\lambda_{YM},T)$ and $f_p (\lambda_{YM},T)$ are interaction terms, non-trivial to calculate \cite{arnold,strickland} but can be neglected within an order-of-magnitude calculation.

At $T<T_c$, the EoS is given by \cite{rafelski}:

\begin{eqnarray}
\label{eq:Pressure_scalar}
p_h(T) & = & \frac{m_h^{2}T^{2}}{2\pi^{2}}K_{2}\left(\frac{m_h}{T}\right) +\sum_{n>4} \order{\left( \frac{T}{N_c \Lambda_{YM}} \right)^n},\\
e_h(T) & = & \frac{m_h^{2}T}{2\pi^{2}}\left[ K_{2}\left(\frac{m_h}{T}\right)T-\frac{m_h}
{2}\left(K_{1}\left(\frac{m_h}{T}\right)+K_{3}\left(\frac{m_h}{T}\right) \right)\right] +\sum_{n>4} \order{\left( \frac{T}{N_c \Lambda_{YM}} \right)^n}
\end{eqnarray}
where $K_1$, $K_2$ and $K_3$ are modified Bessel functions and the summation is of order four.

\subsection{The transport properties \label{transports}}
Around the critical temperature $T_c$, the theory is expected to be strongly coupled.  Thus, the shear viscosity $\eta$ and the relaxation time $\tau_\pi$ should be small. Particularly, from quantum mechanics and string theory arguments \cite{mikloseta,kss}, one expects $\eta/s \sim \order{10^{-2}}$ and $\tau_\pi \sim \eta/(sT)$ \cite{hicexp2}, where $s$ is the entropy density. We also note that $\eta/s$ is generally irrelevant for a nearly homogeneous solution since shear gradients vanish for this solution.  The effect of these anisotropies might however have a relevance for modern dark energy \cite{floerchinger}.

The behaviour of bulk viscosity is however non-trivial and might have crucial phenomenological consequences, both in heavy ion collisions and in cosmology.  From symmetry arguments as well as quantitative calculations, it is generally agreed \cite{amybulk} that $\zeta/s$ is vanishing at $T \gg T_c$. However, at $T \sim T_c$, the system is strongly interacting and it is in the vicinity of a phase transition with formation of condensates. One should remember that while shear viscosity depends on momentum transport, hence on elastic reactions, bulk viscosity is more dependent on diffusion {\em across the diagonal of the energy momentum tensor}, thus, on the thermalization time of inelastic reactions \cite{weinberg,jeon}. Therefore, when condensates form, it is natural to expect bulk viscosity to peak even when shear viscosity is small \cite{mebulk1,mebulk2,cavitation}.

In the $SU(2)$ case, this peak should diverges around the deconfinement second order phase transition  due to the sensitiveness of the Polyakov loop expectation value to $T$ around $T_c$. However, for all $SU(N_c>2)$ cases, the deconfinement transition is of first order and the $T_c$ centered peak does not diverge \cite{kharbulk,gubser,latbulk}. Although its height and width dependence on $N_c$ are not clear, it is reasonable to expect that the peak height goes as the chemical thermalization timescale, which is sensitive to the difference in entropies of the gluonic and hadronic phases during their coexistence period, this is, $N_c^2$.  It is also not clear, from fundamental arguments, how the peak width changes as a function of $N_c$.  Bulk viscosity of a mixture of two phases is additive \cite{jeon}, which should reduce the width below $\sim N_c^2$ scaling.  

For a high enough peak, hydrodynamics solutions such as the Hubble expansion become unstable against small perturbations \cite{mebulk2,cavitation}. This scenario, when viscous forces overwhelm advective forces in the zero chemical potential limit, can be an indication that hydrodynamics fails as an effective theory. However, this is {\em not} the case here, because momentum equilibration happens fast and the lack of chemical equilibrium is related to the presence of a phase transition.  Furthermore, the experimental evidence from heavy ion collisions \cite{hicexp,hicexp2,hicexp3} suggests that matter around $T_c$ continues to behave as a very good fluid independently of the existence of a bulk viscosity peak. 

\begin{figure*}[h]
\epsfig{width=18cm,clip=1,figure=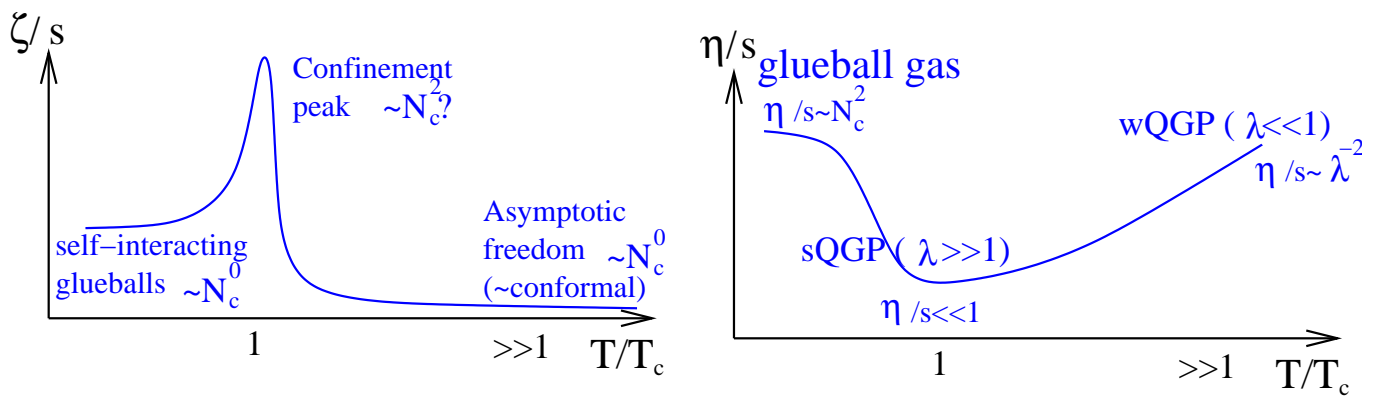}
\caption{\label{etafig} Conjectured behavior of the shear (right panel) and bulk viscosity (left panel) as a function of temperature for Yang-Mills matter}
\end{figure*}

In summary, the likely behavior of the shear and bulk viscosities for $SU(N_c)$ theories is shown in Fig. (\ref{etafig}): bulk viscosity has a Gaussian-like peak around $T_c$, while shear viscosity has a dip in the same region. The Gaussian peak is likely to be very sharp around $T_c$, while the dip goes up logarithmically with temperature \cite{mikloseta}. Below deconfinement, shear viscosity is large, while bulk viscosity is small because the glueballs are infinitely massive and non-interacting. Using the weakly coupled Lagrangian in the previous section (equation \ref{eftlagrangian}), we expect:

\begin{equation}
\frac{\eta}{s} \sim \frac{\Lambda_{YM}}{T^2} \eqcomma \frac{\zeta}{s} \sim \frac{T}{\Lambda_{YM}}
\end{equation}
\section{Cosmology with a hidden $SU(N_c)$}
\subsection{Introduction \label{seccosmo}}
From the discussion of the previous section, we saw that a hidden $SU(N_c)$ sector with $\Lambda_{YM} \geq 1$ TeV could {\em potentially} provide several of the ingredients normally used to construct cosmological models naturally.  If, as argued in \cite{mebulk1,mebulk2}, at $T\sim T_c$ bulk viscosity overwhelms advective pressure but hydrodynamics continues to work, the effective pressure will be negative, as noted before \cite{lima}.

Thus, a peak in bulk viscosity as well as a mixed phase during confinement could naturally provide the large cosmological constant needed for Inflation. Glueballs could be a suitable candidate for dark matter and the small bulk viscosity of glueball's self-interactions could account for dark energy, switching on only recently. All of these features would be dictated by the single scale $\Lambda_{YM}$.

Admittedly, this scenario is very different from standard cosmology \cite{baumann}.
There, the universe starts out cold in a semiclassical field configuration dominated by the vacuum energy, then it finds the true vacuum while converting that energy into thermally distributed particles via reheating. Here, expansion proceeds from an initial Planck temperature and all of the history of the universe is reproduced by changes in the equation of state and transport coefficients. From our vantage point, this scenario could look very similar to the cosmological standard model for all observables.

We note that a similar inflationary model was already explored in \cite{qcdbulk}, where the QCD phase transition would be responsible for a negative pressure driving Inflation. However, as the authors found out, the scales of Inflation and of the QCD deconfinement transition did not match by orders of magnitude, hence the need for a QCD-like beyond the standard model theory.

In this context, during the pre-inflationary era, between temperatures $T_P \sim G^{-1/2}$ and $T\sim \Lambda_{YM}$, the universe would be composed of a hot plasma of $N_C^2$ ``gluons'' of the hidden sector plus standard model matter and radiation. The contribution of the latter to the total entropy density is expected to be subdominant for large $N_c$. As $T \rightarrow T_c$, this plasma would become a good fluid, in local thermal equilibrium. In this regime, thermal fluctuations are approximately Poissonian ($ \frac{1}{e} \frac{de}{dx} \sim T$) and bulk viscosity is negligible.

As $T=T_c$, the bulk viscosity shoots up, so the effective pressure becomes:

\begin{equation} 
\label{zetainfl}
p - \zeta(T \simeq T_c) \frac{\dot{a}}{a} \ll 0
\end{equation}

\noindent{where} $a$ is the cosmological scale factor. Intuition from transport theory would imply that if the effective pressure is negative then the Knudsen number is large, thus, terms beyond shear and bulk viscosity (from Israel-Stewart hydrodynamics \cite{hicexp3}), transport theory and so on should be taken in consideration. However, bulk viscosity diverges due to non-conformal strongly coupled dynamics rather than due to the lengthening of the mean free path. Therefore, at least in an approximately homogeneous universe, we expect those terms to stay small \cite{mebulk1}.

Driven by a negative effective pressure, the universe acquires a cosmological constant, a la \cite{goo,bulkdark1,bulkdark2}. However, differently from those scenarios, this effective pressure gets large, dominating pressure and energy density (eq. \ref{zetainfl}), thus, closely matching the dynamics of the Inflation era until $T \leq T_c$. 

The duration of Inflation depends on the $N_c$ in $SU(N_c)$, since $\zeta(T)$ (Fig.\ref{etafig}) will maintain the peak value of $T_c$ for the whole mixed phase, that is, approximately $N_C^2$ in energy density.  Therefore, one can choose a convenient $N_c$ to achieve an appropriate number of efoldings. As soon as $T <T_c$, the $SU(N_c)$ plasma freezes out into a self-interacting gas of heavy glueballs ($m_g \sim \Lambda_{YM}$) whose entropy content (eq. \ref{eq:Pressure_scalar}) is much smaller than the entropy of the standard model sector and whose bulk viscosity goes as $\zeta/s \sim T/m \ll 1$ \cite{weinberg}. Hence, Inflation naturally stops and the energy-matter content of the dark sector gets transferred to non-relativistic and non-interacting massive glueballs, which compose dark matter.

\subsection{Implementation}
The usual Friedmann equations are simply the continuity equation and Einstein's equation for an isotropic and homogeneous background. As such, the only dynamical value is the scale factor of the universe $a(t)$ and non-gravitational dynamics is provided by the equation of state and the transport coefficients. Numerically, it is usually convenient to write those expressions in conformal time ($\tau$) coordinates rather than locally Minkowski time ($t$) coordinates. The two are related by:

\begin{equation}
    \tau = \int \frac{dt}{a(t)} \eqcomma
    \frac{da}{d \tau}=a'
\end{equation}

In these coordinates, the FRW equations \cite{cosmobook} are: 

\begin{equation}
    a'^2 + k a^2 = 2 \alpha e a^4 
\label{eq1}
\end{equation}

\begin{equation}
    a'' + ka = \alpha \left( e - 3p  \right) a^3
\end{equation}
where $\alpha=8 \pi G/3$ and $G$ is the gravitational constant. Bulk viscosity turns the effective pressure with respect to local time into: 
\begin{equation}
    p_{ef} \rightarrow p - \zeta \frac{1}{a} \frac{da(t)}{dt}
\end{equation}

These equations are usually analytically or semi-analytically solvable for a simple equation of state, such as $p=c_s^2 e$ or at least a polytrope $p = C e^n$. However, for us, as seen in sections \ref{phases} and \ref{transports}, this is not the case, especially considering the effect of bulk viscosity. Nevertheless, these equations can be put into a form amenable to simple numerical integration.

Defining the non-local Hubble constant, one can transform equation \ref{eq1} into an algebraic one:

\begin{equation}
    f=  \frac{a'}{a}\eqcomma a = \sqrt{\frac{f^2+k}{2 \alpha  e}}
\end{equation}

which, after some reshuffling, becomes:

\begin{equation}
    \begin{array}{cl}
    e' =& -3 f (e+p_{ef}) \\
    f' =& \frac{1}{2} \left( \frac{e-3p_{ef}}{e}  \right) \left( f^2 +k \right) - k -f^2\\
    p_{ef} =& p - 3\zeta e^{3/4} \frac{f}{a} +\frac{e}{3}
    \end{array}
\label{friedm}
\end{equation}

These equations can be solved numerically using the equation of state in \ref{phases}, constructed by interpolating lattice data from \cite{panero}, and a bulk viscosity of \ref{transports}, modelled by a Gaussian function, to get the number of efoldings in terms of the parameters of the bulk viscosity peak.

We note that the position of the peak is a free parameter of the theory.  The height and the width, however, should be calculable from first principles, for example from lattice simulations. Nevertheless, in this work, we treat them as free parameters because we want to examine the relationship between these quantities and the inflationary dynamics. Such a study is important in case future lattice calculations determine the precise shape of this peak, in which case our inflationary model may be falsifiable.

\section{Bulk viscosity-driven Inflation}

\begin{figure*}[h]
    \epsfig{width=13cm,clip=1,figure=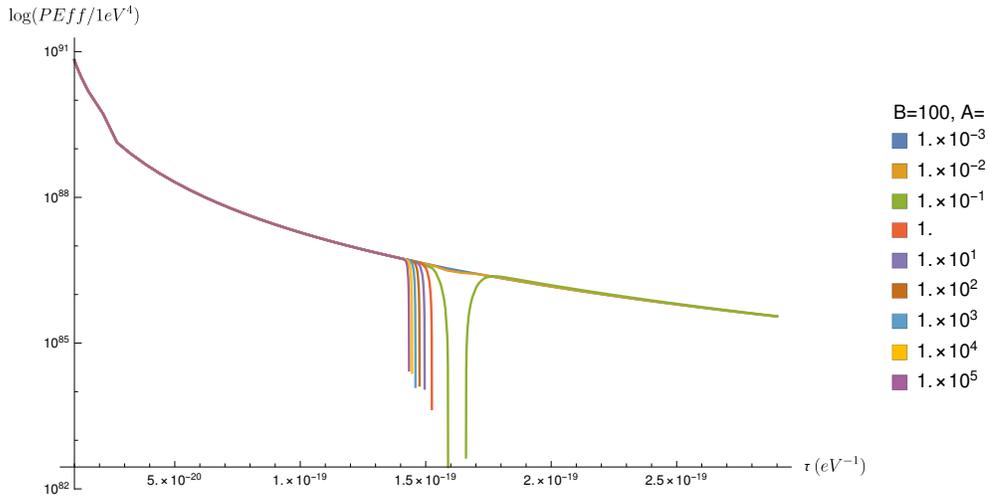}
    \caption{\label{pefftime} The effective pressure as the function of conformal time, for different peak heights of the bulk viscosity Gaussian function.}
\end{figure*}

\begin{figure*}[h]
\epsfig{width=13cm,clip=1,figure=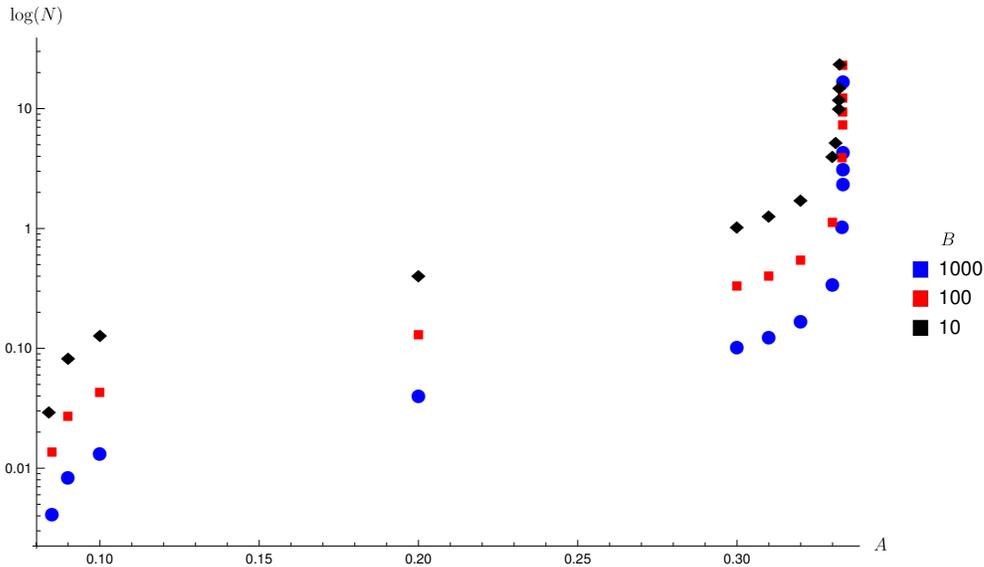}
\caption{\label{efolds} The number of efoldings as a function of the height of the peak in bulk viscosity, for several peak widths as a percentage of the height }
\end{figure*}

The results of the numerical calculations, which formed the bulk of the computational work in \cite{thesis}, are summarized in Fig. \ref{pefftime} and Fig. \ref{efolds}.   As Fig. \ref{pefftime} shows, the time evolution of the effective pressure follows the pattern summarized in section \ref{seccosmo}. At the approach of the bulk viscosity peak the effective pressure becomes negative, stays negative for an amount of time monotonically dependent on the peak's height, and then resumes expansion, returning to be positive.  

However, for values below the critical value of height $A \simeq 0.37$, the universe enters a never-ending Inflation phase, which is quite different from the ``eternal Inflation'' of \cite{linde} and, unlike it, incompatible with the present universe.

We define the number of efoldings $N$ as:
\begin{equation}
    N=\int^{t_f}_{t_i} \frac{\dot{a}}{a}dt = \int^{\tau_f}_{\tau_i} \frac{a'}{a}d\tau
\end{equation}
To compare existing limits on $N$ from standard model inflationary theory, we numerically find the duration of the bulk viscosity peak in conformal time and calculate it for this period. The result confirms that the number of efoldings depends in a diverging way on the peak's height $A$. Initially, $N$ increases approximately exponentially with $A$ but, as this quantity approaches a critical value, an arbitrary number of efoldings can be obtained, converging to a never-ending Inflation.

The existence of this phase, which mirrors the original problems with the latent heat-driven Inflation of \cite{guth}, is actually not so surprising. We remember that the physical action of viscosity is to convert ``work'' (the expansion of the universe) into ``heat'' (entropy density) and the rate of this conversion is proportional to $\zeta (\partial s)^2$ \cite{weinberg}.  We also note that, in a curved spacetime, energy is conserved only ``locally'', while the FRW equations control global dynamics. Hence, if homogeneity is imposed, it is not surprising that, for a high enough peak entropy, energy creation is stronger than the exponential expansion of the universe, triggering an Inflation epoch that never ends.

The appearance of a hot eternal Inflation, however, diminishes the naturalness appeal of our model, especially since the height of the peak is only a free parameter due to our ignorance. It remains to be seen how natural it is the obtaining of finite Inflation with respect to the expected height of the $\zeta/s$ peak.  

The dynamics described here is only weakly dependent on the width of the peak, as Fig. \ref{efolds} shows. However, it strongly depends on the relative location of $T_c$ with respect to the Planck scale, the essential reason why the QCD based model of \cite{qcdbulk} did not work.

\begin{figure*}[h]
  \epsfig{width=13cm,clip=1,figure=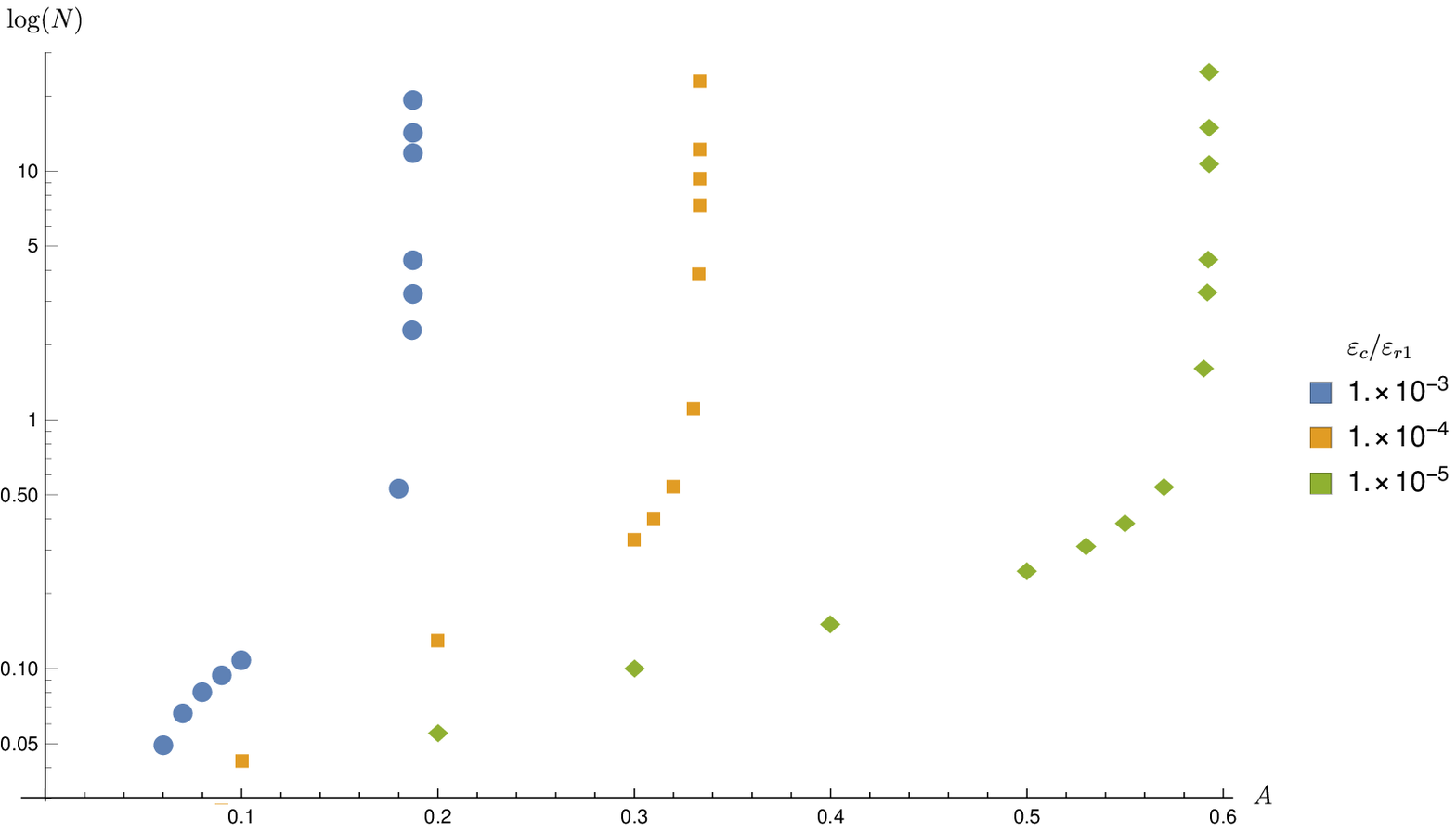}
  \epsfig{width=13cm,clip=1,figure=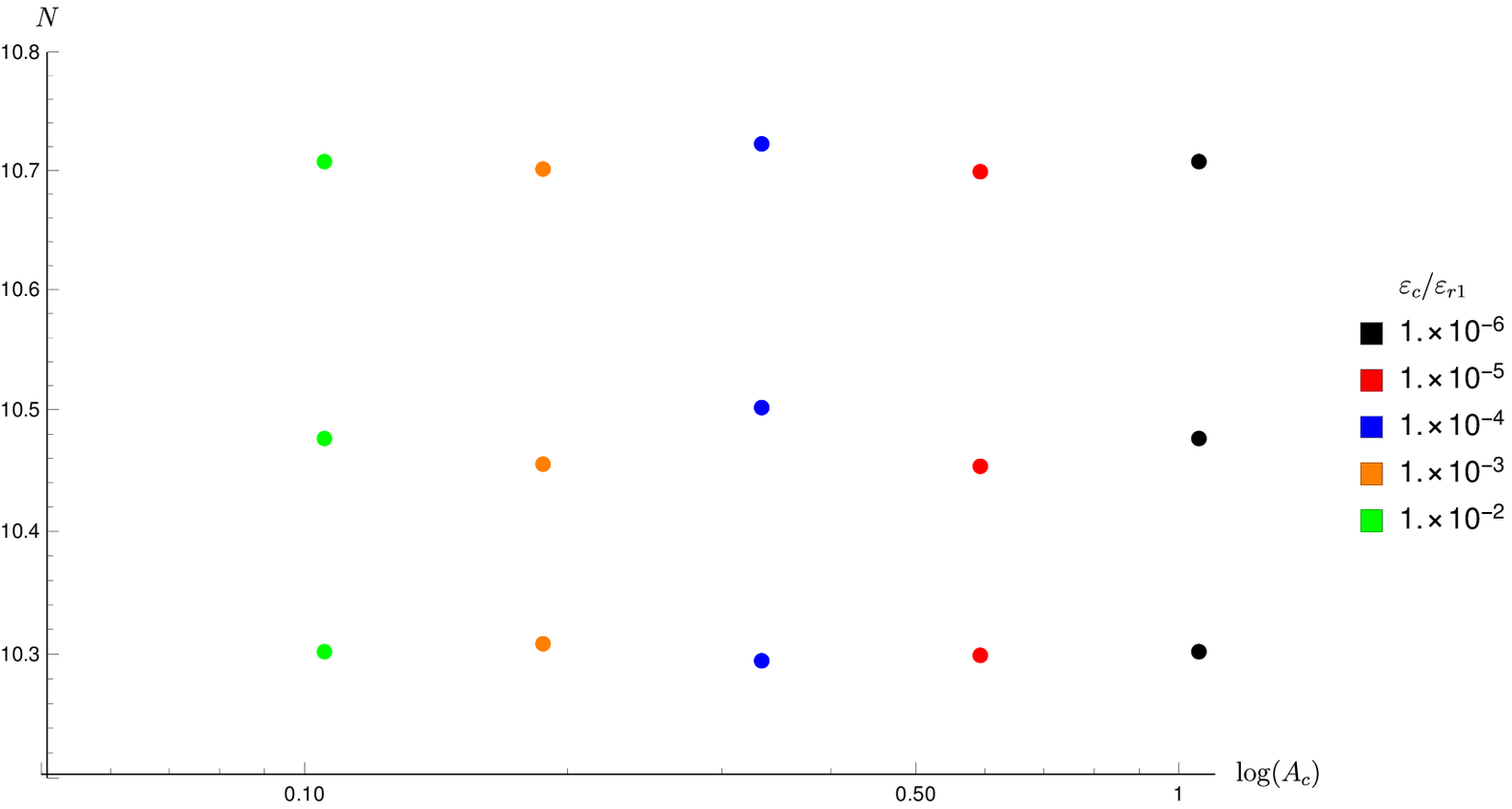}
\caption{\label{efoldsec} The number of efoldings as a function of the height of the peak in bulk viscosity for several peak positions as a fraction of the planck energy density.
The top figure represents a zoomed-in version of the bottom one  }
\end{figure*}

To study this dependence, the only free parameter of the model is $e_c$, since $T_c \sim \Lambda_{YM}$ and the equation of state has been rewritten such that $p(e)$, turning its dependence with temperature implicit. We varied $e_c/e_{planck} \sim e_c G^2$ in the interval $[10^{-5},1]$, where, qualitatively, the equivalent curves to Fig. \ref{pefftime}  look similar, differing only on a shift in the position of the bulk viscosity peak, this is, the conformal time at which the effective pressure gets negative. This different location affects the number of efoldings approximately linearly, as shown in Fig. \ref{efoldsec} in a logarithmic scale. Note that its shape is qualitatively similar to that of Fig. \ref{efolds}.
 
In the continuum limit of $N_c$, under the 't 'Hooft scaling, and assuming that the peak height is proportional to $N_c^2$, one can conclude that the number of efoldings $N \geq \order{10}$ implies a constraint on $N_c$ and $\Lambda_{YM}$. This constraint together with the calculation of the abundance of dark matter (i.e. glueballs) should make our theory falsifiable. This will be discussed in the next section.

\section{Discussion, challenges and prospects}
Given a successful evasion of the never-ending Inflation issue outlined in the previous section, the next phenomenological challenge for our model would be a successful description of the current dark matter abundance.

From the Lagrangian in Eq. \ref{eftlagrangian} it is clear that the interaction of dark matter particles is local on a scale of $\Lambda_{YM}$ and, in the 't Hooft limit, it gets suppressed by $N_c^{-2}$. At $T<T_c$, the glueballs become weakly self-interacting and come out of equilibrium, so one cannot trust the calculation of the equation of state in the previous section from the end of deconfinement onwards.

However, one can functionally think of the gas of glueballs as a ``dust'' of conserved particles, this is, a distribution of non-relativistic and non-interacting massive particles. Assuming an ideal equation of state for standard model matter (energy density $e_S$, pressure $p_S=e_S/3$, degeneracy $g_S$), the equations \ref{friedm} of the last section can be appended by:
\begin{equation}
    e = m_h n + e_S \eqcomma p = \frac{e_S^{1/4}}{g_S m_h} n + \frac{1}{3}e_S \eqcomma e_S \sim g_S T^4
\end{equation}
Where $n$ for glueballs' number density. We also add a conservation equation for the number density of glueballs:
\begin{equation}
    \frac{d n}{d \tau}= f n + \order{\frac{1}{N_c^2}\left( \frac{T}{\Lambda_{YM}} \right)^{n\geq 4}}
\end{equation}

\noindent{where the last} term vanishes rapidly after the deconfinement phase transition, since  we do not expect scattering and annihillation of glueballs to be sizeable in the confined phase.

We note that dark matter density at its formation should increase monotonically, $n(T=T_c)\sim N_c^2 \Lambda_{YM}^3$, since dark matter mass density should be comparable to energy density at deconfinement.  In the previous section, however, we saw that, in Inflation, $N_c$ and $\Lambda_{YM}$ are also correlated for a fixed number of efoldings.  Our model, therefore, predicts a correlation between the number of efoldings and the dark matter abundance which is in principle testable. These equations will be examined in a forthcoming publication.

The gas of glueballs naturally tracks the perturbations that formed in the inflationary era. While the usual source of perturbations - quantum fluctuations of the inflaton field \cite{baumann} - are inapplicable here, hydrodynamic instabilities could provide an alternative mechanism for generating fluctuations, since it has been shown \cite{mebulk2} that the Hubble hydrodynamic solution is unstable against small perturbations. Furthermore, provided that the scale separation between the macroscopic Hubble factor $\dot{a}/a$ and the microscopic bulk viscosity $\zeta/(sT)$ is wide enough, these perturbations could have a scale-free spectrum, based on  Kolmogorov's arguments \cite{turbulence}.

Since the glueballs are heavy, weakly self-interacting and, by assumption, flavourless, its gas naturally plays the role of the sinks of cold dark matter assumed in $\Lambda CDM$ cosmology. 
To complete our model's scenario, we mention that a residual interaction between glueballs might have a role in the constitution of dark energy explaining why it switched on only recently. For a weakly coupled massive gas, $\eta/s \gg 1$, but the relaxation time is also large, $\eta/(Ts) \gg \dot{a}/a$.  
Hence, the shear viscosity will have a large turn-on time and it will only appear long after the formation of the glueballs. Quantitatively, this could be implemented by solving the FRW equations with Israel-Stewart dynamics \cite{hicexp3} 
and adding the Israel-Stewart equation for $\Pi$:
\begin{equation}
p_\zeta \rightarrow p- \Pi \eqcomma \tau_\pi \left( \dot{\Pi} + \Pi \frac{\dot{a}}{a} \right) + \Pi = 3 \zeta \frac{\dot{a}}{a}
\end{equation}

\noindent{with} $\tau_\pi \sim \eta/(Ts) \sim \Lambda_{YM} N_C^2/T$. In this context, large enough $\Lambda_{YM}$ and $N_c$ might explain a late switch on of dark energy. Finally, we should mention that a non-zero $\theta$ parameter \cite{theta} associated with the dark sector could in principle account for baryogenesis. It would lead to Eq. \ref{lagrangian} being augmented by a term of the form $\sim \theta Tr_a\left[ \epsilon_{\alpha \beta \mu \nu} F_a^{\alpha \beta} F_a^{\mu \nu} \right]$ and a suppressed interaction between the standard model and the dark sector. 

In conclusion, in this work we have argued that some ideas developed in quark-gluon plasma physics when extended to a hypothetical Yang-Mills theory without flavours and with a transition scale in the \textit{TeV} range, can unify apparently unrelated features of the standard cosmological model such as Inflation and dark matter.
We have shown that Inflation due to the bulk viscosity peak at deconfinement can reproduce the required number of efoldings. Nevertheless, avoiding a hot never-ending Inflation phase might require some fine-tuning. We have listed the cases capable of falsifying our model and the potential ways this scenario could attempt to explain the standard features of our universe: dark matter, perturbations, dark energy and baryogenesis.
We hope quantitative progress on the points listed above will confirm whether this scenario is phenomenologically viable.

\vskip 0.3cm
GT acknowledges support from FAPESP proc. 2017/06508-7 and CNPQ bolsa de produtividade 301996/2014-8. MM acknowledges support from CNPQ process 132657/2015-5, Global Affairs Canada through DFATD, by the concession
of an ELAP scholarship at McGill University and CRC-TR211 by the concession of a Master's qualification fellowship at J. W. Goethe University.  We thank Marco Panero, Jose Ademir Lima, Rodolfo Valentim, Pedro Holanda and Jean-Sebastian Gagnon for discussions and suggestions.

\end{document}